\documentclass[a4paper,10pt]{amsart}

\usepackage{amsmath,amssymb,amsthm,a4wide}

\usepackage{tikz}
\usetikzlibrary{shapes}
\usetikzlibrary{intersections}

\usepackage{graphicx}
\usepackage{graphics}
\usepackage[foot]{amsaddr}

\newtheorem{theorem}{Theorem}

\DeclareMathOperator*{\argmax}{arg\,max}

\begin{document}

\title[]{Stability, Fairness and random walks \\in the bargaining problem}

\author{Jakob Kapeller}
\address{(J.K.) Institute for Comprehensive Analysis of the Economy and Department of Economics, Johannes Kepler University Linz, Altenbergerstrasse 69, Linz, Austria}
\email{jakob.kapeller@jku.at}

\author{Stefan Steinerberger}
\address{(corresponding author: S.S.) Department of Mathematics,  Yale University, 10 Hillhouse Avenue, New Haven, CT 06511, United States of America}
\email{stefan.steinerberger@yale.edu}

\begin{abstract} 
We study the classical bargaining problem and its two canonical solutions, (\textsc{Nash} and \textsc{Kalai-Smorodinsky}), from a novel point of view: we ask for stability of the solution if both players are able distort the underlying bargaining process by reference to a third party (e.g. a court). By exploring the simplest case, where decisions of the third party are made randomly we obtain a stable solution, where players do not have any incentive to refer to such a third party. While neither the Nash nor the Kalai-Smorodinsky solution are able to ensure stability in case reference to a third party is possible, we found that the Kalai-Smorodinsky solution seems to always dominate the stable allocation which constitutes novel support in favor of the latter.
\end{abstract}
\maketitle

\section{Introduction}
\subsection{The bargaining problem.} The bargaining problem tries to capture the difficulty of 'fair division' and is one of the cornerstones of cooperative 
game theory. Two players 
have to agree on a division of an object (say, one piece of cake) that leads to them receiving utilities $U_1$ and $U_2$: it is in their interest 
to come to an agreement as otherwise both get a predefined amount $c = (c_1, c_2)$ which is strictly dominated by other possible payoffs (i.e. $c$ is not pareto-optimal).
An additional difficulty is that the players are not indistinguishable but have different utility functions (one of them is more hungry than the other). The challenge is to understand the properties
a fair solution should exhibit and then see which solutions satisfy all these desired properties.
\begin{figure}[h!]
\begin{tikzpicture}[scale = 1]
\draw [ultra thick, ->] (0,0) -- (2.5,0);
\draw [ultra thick, ->] (0,0) -- (0,2.5);
\draw [dashed] (0,0) -- (2,2);
\filldraw (0,0) circle (2pt);
\node at (-0.1,-0.3) {\Large $c$};
\node at (2.7,-0.4) {$U_1$};
\node at (-0.4,2.2) {$U_2$};
\draw[ultra thick] (0,2) to[out=0,in=120] (1.6,1.3) to [out=300,in=100] (2,0);
\end{tikzpicture}
\vspace{-10pt}
\caption{The feasibility set $\mathcal{F}$ is the convex set bounded by the axes and the curve. Note that the feasibility set is \textit{not} symmetric
around the $45^{\circ}$ slope.}
\end{figure}
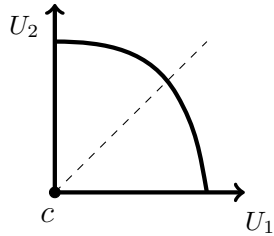

Let us formally introduce the feasibility set as a compact set $\mathcal{F} \subset \mathbb{R}^2$ that contains all payoff vectors which can be realized by
the two players: if $(x, y) \in \mathcal{F}$, then it is possible for the players to agree to an outcome of the bargaining problem where Player 1 receives utility $U_1$
and Player 2 receives utility $U_2$. Furthermore, we will introduce the disagreement point $c \in \mathbb{R}^2$ as the
outcome that is being realized if the players do not agree on an outcome. The tuple $(\mathcal{F}, c)$ constitutes a bargaining
problem. We will follow the typical convention of assuming that $\mathcal{F} \cup \left\{c\right\}$ is a convex set: any linear combination of elements may be realized if the game is played iteratively as the players are capable of forming binding contracts between them.
This paper asks for the relative performance of the two main canonical solutions of the bargaining problem -- the Nash solution and the Kalai-Smorodinsky solution -- with respect to the stability of bargaining outcomes.  Specifically, we analyze the case, where both players may dispute the assignment of a certain item subject to the bargaining problem by referencing a third party, which decides randomly on the assignment of said item. We think this case is of a central importance as it resembles typical features of real-world bargaining problems, especially the possibility of sectoral conflict, i.e. situations where a partial compromise is attained, while some specific items remain heavily contested.

This paper proceeds as follows: First we introduce the canonical solutions to the bargaining problem in Section 2, before exploring the possibility of partial compromise and sectoral conflict in bargaining in Section 3. Section 4 provides a generalization and formalization of our conceptual argument and introduces our core theorem on the nature of partial compromise in bargaining. Section 5 shows the detailed proofs for our results, while Section 6 concludes the paper by mapping possible extensions of our approach.

\section{Canonical solutions to the bargaining problem.} A \textit{solution} $\mathcal{S}$ of the bargaining problem is defined as a mapping $\mathcal{S}: (\mathcal{F}, c) \rightarrow \mathcal{F}$ 
that assigns to each bargaining problem $(\mathcal{F}, c)$ a unique payoff vector $\mathcal{S}(\mathcal{F}, c) \in \mathcal{F}$. The approach taken in the
highly influental paper of Nash \cite{nash} is the \textit{axiomatic method}, according to which it is better  to list properties of acceptable solutions instead of defending some specific outcome. This approach simplifies the discussion of various different solutions
by focusing on their performance relative to the proposed axioms.
The first three of the properties proposed by Nash are exceedingly natural and have essentially gone unchallenged. All of them are also satisfied by
the Kalai-Smorodinsky solution; we also accept them as minimal requirements for any reasonable solution in what follows.\\

\underline{Axiom 1.} \textsc{ Symmetry.} If the feasibility set $\mathbb{F}$ is symmetric in the sense that 
$$(x,y) \in \mathcal{F} \implies (y,x) \in \mathcal{F}$$
and if the disagreement point is symmetric ($c = (c_1, c_2)$ satisfies $c_1 = c_2$), then the bargaining problem does not distinguish
between Player 1 and Player 2. It is only natural that for a solution $\mathcal{S}$ to be considered fair, it should assign them an equal share of the good, i.e. $\mathcal{S}(\mathcal{F}, c) = (x,x)$ for some $x \in \mathbb{R}$.\\

\underline{Axiom 2.} \textsc{ Invariance under an affine change of coordinate.} The solution should be well behaved under the application of some linear 
transformation $T: \mathbb{R}^2 \rightarrow \mathbb{R}^2$ given by 
$$ T(x,y) = (a_1 x + b_1, a_2 y + b_2) \qquad \mbox{with} \quad a_1, a_2 > 0 \quad \mbox{and} \quad b_1, b_2 \in \mathbb{R}.$$
Axiom requires a solution $\mathcal{S}$ to satisfy
$$ \mathcal{S}(T\mathcal{F}, Tc) = T\mathcal{S}(\mathcal{F}, c).$$
If two players were to divide a cake between them, it should not
be of importance whether they measure the cake in kilograms or pounds; most importantly, if one player were to switch his scale from
kilograms to pounds, it should have no implications whatsoever on the solution.\\

\underline{Axiom 3.} \textsc{ Pareto optimality.} Pareto optimality requires that a fair solution should not be strictly dominated by any
other element in the feasibility set. If $\mathcal{S}(\mathcal{F}, c) = (x,y)$, then 
$$ \mbox{for every} \quad (z,w) \in \mathcal{F} \quad \mbox{either} \quad z \leq x \quad \mbox{or} \quad w \leq y.$$
If this were violated, then there would be another possible division of the good where at least one player does better but none of
them worse.
Players, if dividing a cake between them, should not throw away any part of the cake.

\begin{figure}[h!]
\begin{tikzpicture}[scale = 1]
\draw [ultra thick, ->] (0,0) -- (3,0);
\draw [ultra thick, ->] (0,0) -- (0,3);
\filldraw (1,1) circle (2pt);
\node at (1.1,0.7) {$c$};
\node at (2.7,-0.4) {$U_1$};
\node at (-0.4,2.2) {$U_2$};
\draw[ultra thick] (0,2) to[out=0,in=160] (1.1,1.8) to [out=330,in=100] (2,0);

\draw [ultra thick, ->] (4,0) -- (7,0);
\draw [ultra thick, ->] (4,0) -- (4,3);
\filldraw (5,1) circle (2pt);
\node at (5.1,0.7) {$c$};
\node at (4+2.7,-0.4) {$U_1$};
\node at (4-0.4,2.2) {$U_2$};
\draw  (5,1) -- (5,1.8);
\draw  (5,1) -- (5.76,1);
\draw (4,2) to[out=0,in=160] (5.1,1.8) to [out=330,in=100] (6,0);
\draw[ultra thick] (5,1.82) to [out=330,in=100] (5.76,1);
\end{tikzpicture}
\caption{Pareto-optimal points with a payoff dominating the disagreement point.}
\end{figure}
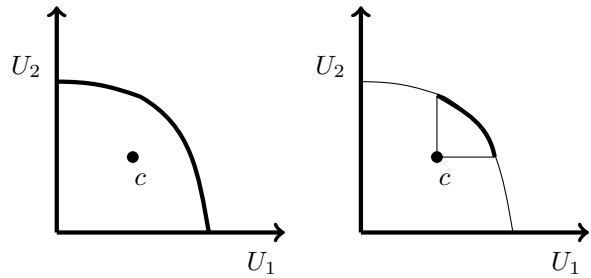

Sometimes it is explicitely required as axiom (\textsc{Individual Rationality}) that any solution assigns to each player at least as high a payoff as assigned by the
disagreement point $c$. Clearly, any solution violating that condition is a bad predictor of the outcome of the game because the
player would simply refuse to come to an agreement. We will always assume that condition throughout the paper.

\subsection{The Nash solution.} Nash's solution \cite{nash} is based on the following additional fourth axiom which he considers 
natural for a solution to satisfy. 
\begin{quote}
\underline{Axiom 4.} \textsc{Independence of irrelevant alternatives (IIA).}  Given two bargaining problems with the same disagreement point $(\mathcal{F}_1, c)$
and $(\mathcal{F}_2, c)$, this axiom demands that if $\mathcal{F}_1 \subset \mathcal{F}_2$ and 
$$ \mathcal{S}(\mathcal{F}_2, c) \subset \mathcal{F}_1 \quad \mbox{then necessarily also} \quad \mathcal{S}(\mathcal{F}_1, c)  = \mathcal{S}(\mathcal{F}_2, c) .$$
\end{quote}
The property is easily explained: suppose a solution decides on a payoff for a particular bargaining problem, then the removal of a subset of the feasibility set not containing
the proposed payoff should not affect the solution. Alternatives that are not being considered a 'fair' payoff anyway should not affect the final outcome.
Nash proved that there exists a unique solution $\mathcal{S}_{Nash}$ satisfying Axioms 1,2,3,4. The Nash bargaining solution $\mathcal{S}_{Nash}$ is given by maximizing
the Nash product
$$\mathcal{S}_{Nash}(\mathcal{F}, c) = \argmax_{(x,y) \in \mathcal{F}}{(x-c_1)(y-c_2)}.$$
Axiom 4, often abbreviated as \textsc{IIA}, has been the focus of criticism as early as in the 1957 book of Luce \& Raiffa \cite{luce}. As pointed out in Peters \& Wakker \cite{pe}, while
\textsc{IIA} is essentially responsible for computational simplicity of the solution, there seems to be no powerful argument
in favor of Axiom 4. Moreover, computational simplicity also seems to underlie the relative success and popularity of the Nash solution.
'In spite of the criticisms levelled against Nash’s axiom system, his solution
has met with what can only be described an extraordinary success in economics,
although it is probably the case that the profession at large did not
adopt it for its axiomatic foundations, but rather for the ease with which it
can be calculated' (Thomson \cite{thom}).

\subsection{The Kalai-Smorodinsky solution.} A particular idiosyncracy of the Nash solution was pointed out by Kalai \& Smorodinsky \cite{ks}: it is not difficult to construct two bargaining problems with the same disagreement point $(\mathcal{F}_1, c)$, $(\mathcal{F}_2, c)$ and $\mathcal{F}_1 \subset \mathcal{F}_2$, where the pay-off for one player strictly decreases when switching from the first to the second problem, although the underlying set of possible outcomes has been extended. 

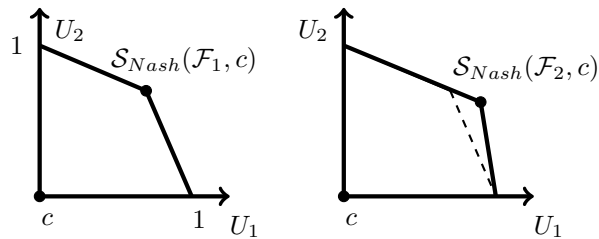
\begin{figure}[h!]
\begin{tikzpicture}[scale = 1]
\draw [ultra thick, ->] (0,0) -- (2.5,0);
\draw [ultra thick, ->] (0,0) -- (0,2.5);
\draw [ultra thick] (2,0) -- (1.4,1.4);
\draw [ultra thick] (1.4,1.4) -- (0,2);
\node at (2.7,-0.4) {$U_1$};
\node at (0.4,2.2) {$U_2$};
\filldraw (0,0) circle (2pt);
\node at (0.1,-0.3) {$c$};
\node at (2.1,-0.3) {$1$};
\node at (-0.3,2) {$1$};
\filldraw (1.4,1.4) circle (2pt);
\node at (1.9,1.8) {$ \mathcal{S}_{Nash}(\mathcal{F}_1, c)$};

\draw [ultra thick, ->] (4,0) -- (6.5,0);
\draw [ultra thick, ->] (4,0) -- (4,2.5);
\node at (4+2.7,-0.4) {$U_1$};
\node at (4-0.4,2.2) {$U_2$};
\draw [thick] (4,2) -- (5.4,1.4);
\draw [thick, dashed] (5.4,1.4) -- (6,0);
\draw [ultra thick] (4,2) -- (5.8,1.25);
\draw [ultra thick] (5.8,1.25) -- (6,0);
\filldraw (4,0) circle (2pt);
\node at (4.1,-0.3) {$c$};

\filldraw (5.8,1.25) circle (2pt);
\node at (6.4,1.7) {$ \mathcal{S}_{Nash}(\mathcal{F}_2, c)$};

\end{tikzpicture}
\caption{Two bargaining problems and their Nash solutions: one has a strictly larger feasibility set but still assigns a smaller outcome to $U_2$. Assuming a unit scale, the payoffs given by the Nash bargaining solution are $ \mathcal{S}_{Nash}(\mathcal{F}_1, c) = (0.7, 0.7)$ and  $ \mathcal{S}_{Nash}(\mathcal{F}_2, c) = (0.8, 0.65)$.}
\end{figure}

In order to remedy this idiosyncratic feature of the Nash solution,
Kalai \& Smorodinsky proposed to substitute the Axiom IV by another one that is characterized by having the property that
such a scenario may never occur.
\begin{quote}
\underline{Axiom 5.} \textsc{Individual Monotonicity.}  Given two bargaining problems with the same disagreement point $(\mathcal{F}_1, c)$
and $(\mathcal{F}_2, c)$, such that for every possible payoff for Player 1 in $\mathcal{F}_1$ the largest possible payoff for Player 2
increases when switching from $\mathcal{F}_1$ to $\mathcal{F}_2$, then the payoff assigned to Player 2 in $\mathcal{F}_2$ should be
at least as large as the payoff assigned to Player 2 in $\mathcal{F}_1$. 
\end{quote}
\begin{figure}[h!]
\begin{tikzpicture}[scale = 1]
\draw [ultra thick, ->] (0,0) -- (3,0);
\draw [ultra thick, ->] (0,0) -- (0,3);
\draw [ultra thick] (2,0) -- (1.4,1.4);
\draw [ultra thick] (1.4,1.4) -- (0,2);
\node at (2.7,-0.4) {$U_1$};
\node at (-0.4,2.2) {$U_2$};
\draw [thick] (2,0) -- (2,2);
\draw [thick] (0,2) -- (2,2);
\draw [thick] (0,0) -- (2,2);
\filldraw (0,0) circle (2pt);
\node at (0.1,-0.3) {$c$};
\filldraw (1.4,1.4) circle (2pt);
\draw [ultra thick, ->] (4,0) -- (7,0);
\draw [ultra thick, ->] (4,0) -- (4,3);
\node at (4+2.7,-0.4) {$U_1$};
\node at (4-0.4,2.2) {$U_2$};
\draw [ultra thick] (4,2) -- (5.8,1.3);
\draw [ultra thick] (5.8,1.3) -- (6,0);
\draw [thick, dashed] (5.4,1.4) -- (6,0);
\draw [thick] (4,2) -- (6,2);
\draw [thick] (6,0) -- (6,2);
\draw [thick] (4,0) -- (6,2);
\filldraw (4,0) circle (2pt);
\node at (4.1,-0.3) {$c$};
\filldraw (5.45,1.45) circle (2pt);
\end{tikzpicture}
\caption{A graphical construction of the Kalai-Smorodinsky solution and its behavior under enlarging the feasibility set in one direction. The bargaining
solution is $\mathcal{S}_{KS}(\mathcal{F}_1, c) = \mathcal{S}_{KS}(\mathcal{F}_2, c) = (0.7, 0.7)$.}
\end{figure}
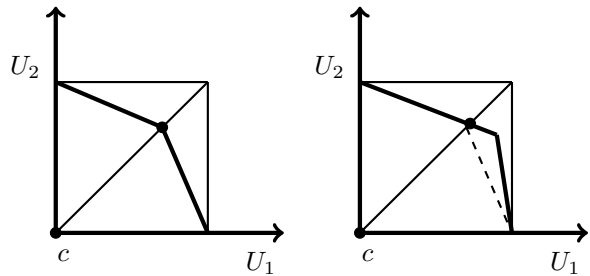
They prove that the only solution satisfying Axioms 1, 2, 3 and 5 is given by the \textit{Kalai-Smorodinsky solution}
$\mathcal{S}_{KS}$, which can be constructed by finding the intersection of the boundary $\partial \mathcal{F}$ of the feasibility set with the line
$$ \mbox{connecting the disagreement point} ~~ c ~~ \mbox{and} \quad \left(\max_{(x,y) \in \mathcal{F}}{x}, \max_{(x,y) \in \mathcal{F}}{y}\right).$$ 
Hence, the property of monotonicity (and the lack thereof in the solution of Nash) establishes the Kalai-Smorodinsky
solution as a natural alternative to the approach of Nash (Thomson \cite{thom}).

\subsection{Other solutions.} A long list of alternative solutions have been proposed; of great importance is the \textit{egalitarian solution}, which can be
regarded as a scale-invariant form of the Kalai-Smorodinsky solution. Chun \cite{ch} proposed the \textit{equal-loss} solution as the point, where the
payoff of both players is equally far removed from their respective ideal outcome. This solution is a special case $(p = \infty)$ of the class of solutions proposed by
Yu \cite{yu} minimizing the $\ell^p-$distance to the ideal point. A particularly geometric approach is taken by the \textit{equal area solution} aiming to compare areas (see Anbarci \& Bigelow \cite{anb}).
There are also dynamical models, most notably the 1953 model of Nash \cite{nash2} yielding again the Nash solution in a different context
or the model of Anbar \& Kalai \cite{anbar}. Another dynamical solution based on comparing the relative marginal gains was proposed by
Perles \& Maschler \cite{per}.

\section{Stability and partial compromise in bargaining}

\begin{quote}
[...] the detailed process of bargaining will differ so widely from one case to
another that any useful theory of bargaining must involve some attempts to
distil out some simple principles which will hold over a wide range of possible
processes. [...] This motivates the idea of looking at some
example(s) of non-cooperative games which correspond to a particular process. (Sutton, \cite{sutton})
\end{quote}

We believe that the mental picture of two agents agreeing on a proper division of a monetary unit or a cake can be misleadingly simple: for instance, to argue for a solution that guarantees a payoff of 0.575 instead of 0.57, in some way assumes an homogeneity of the underlying cake. On the contrary, many real-world bargaining situations involve a diversity of items, which are not necessarily infinitely divisible. This argument even applies to simple examples of bargaining contexts, like dividing a plate of food, drawing borders after an armed conflict or settling relations in the context of divorce. In what follows, we argue that in such contexts actors will practically often reach a partial compromise, where a consensual bargaining solution can be achieved for most, but not all, items under consideration. In order to capture this notion, we explore a model, where individuals may refer to a third party (e.g. a court) to decide the fate of especially contested items randomly. In this setup we look for and explore solutions that are characterized by the property of being stable in the sense that no player has any incentive to refer to such a third party. 

\subsection{Partial compromise and sectoral conflict.} 
To illustrate our argument, we take the problem of finding a fair divorce settlement by employing a bargaining solution $\mathcal{S}$ as an example. If we picture the situation as not amicable and emotions as running high, it is not a stretch to imagine a dialogue along the lines of ''Let's apply the bargaining solution to our joint property except for that vase. You have never liked that vase anyway.'' -- ''Are you kidding me? I have had that vase since college!''. Suppose now that both players have the chance to force the issue and get a court order deciding whose vase it really is: in such a case we can observe the emergence of partial compromise ('Let's apply the bargaining solution...') and -- corresponding -- sectoral conflict ('I've had this vase since college...'), which might provide an incentive to remedy the matter by reference to a third party (a court).
In the simplest form this reference to a third party in case of sectoral conflict can be modelled as free and neutral, i.e. it comes without cost,   decides on assignments at random and has no implication on the remainder of the property. After having settled the issue of the vase, both players agree to use the bargaining solution $\mathcal{S}$.\\

This process of forcing the tiny issue of the vase does not favor any player: either player will get the vase in 50\% of all cases. However, it has the effect of changing the set to which the bargaining solution $\mathcal{S}$ is actually applied. We believe that it is a desirable property of a solution $\mathcal{S}$ to be stable in case of a partial compromise and corresponding sectoral conflict. In this context, stability requires a bargaining solution, where no player has an incentive to initiate the random perturbations associated with referencing a third party. It is instructive to imagine a solution $\mathcal{S}$ \textit{without} that property in practice. Then,
having full knowledge of the solution $\mathcal{S}$, it would actually be in the interest of one the players to \textit{obstruct} the actual decision procedure and instead \textit{delay} the process by repeatedly appealing to the courts for strategic reasons (even though the decision of the court is randomized and might go either way). 

\subsection{A formalization.}
The problem resulting from the above scenario can beformalized in the framework of the bargaining problem by giving players the opportunity have the disagreement point $c$ moved a tiny bit in a random direction, that cannot be anticipated by either player. Given a bargaining problem $(\mathcal{F}, c)$, we fix a very small number $\varepsilon > 0$. Hence, whenever one player forces the issue to court, the court picks a random variable $\theta$, where $\theta$ is uniformly distributed on $[0,2\pi]$ and replaces their original bargaining problem
$$ (\mathcal{F}, c) \quad \mbox{by the new problem} \quad (\mathcal{F}, c + \varepsilon(\cos{\theta}, \sin{\theta})) \quad
\mbox{whenever} \quad c + \varepsilon(\cos{\theta}, \sin{\theta}) \in \mathcal{F}.$$ 
A priori, there is no canonical choice of $\varepsilon$ and we will therefore only require invariance of the solution in
the limit $\varepsilon \rightarrow 0$ (however, this will actually be enough to deduce that the property holds for all
$\varepsilon>0$ for which the statement is well-defined).
\begin{figure}[h!]
\begin{tikzpicture}[scale = 1]
\draw [ultra thick, ->] (0,0) -- (3,0);
\draw [ultra thick, ->] (0,0) -- (0,3);
\filldraw (0.5,0.5) circle (2pt);
\node at (0.5-0.1,0.5-0.3) {$c$};
\node at (2.3,-0.3) {$U_1$};
\node at (-0.3,2.3) {$U_2$};
\draw[ultra thick] (0,2) to[out=0,in=160] (1.1,1.8) to [out=330,in=100] (2,0);

\draw [ultra thick, ->] (4,0) -- (7,0);
\draw [ultra thick, ->] (4,0) -- (4,3);
\filldraw (4.3,1) circle (2pt);
\node at (4.4,0.7) {$c$};
\node at (3.7,2.3) {$U_2$};
\node at (6.3,-0.3) {$U_1$};
\draw [ultra thick] (4,2) to[out=0,in=160] (5.1,1.8) to [out=330,in=100] (6,0);
\end{tikzpicture}
\caption{A bargaining problem and the new problem after having the disagreement point moved in a random direction: in this situation, the outcome of the process has favored
Player 2.}
\end{figure}
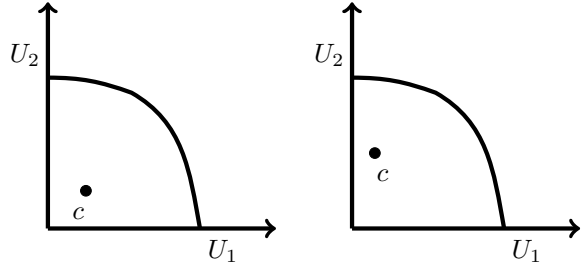
Let us now formally state the axiom. We use the symbol $\mathbb{E}$ to denote the expectation with respect to the uniform distribution of the random variable $\theta$. 
\begin{quote}
\underline{Axiom 6}. \textsc{Invariance in case of sectoral conflict (ISC).} We say that a solution $\mathcal{S}$ is invariant to sectoral conflicts if for every bargaining problem $(\mathcal{F}, c)$ and every $i= 1,2$
$$\forall f \in \mathcal{F} ~~~\qquad~\lim_{\varepsilon \rightarrow 0}{\quad \mathbb{E}_{} \frac{1}{\varepsilon^2}\left[\mathcal{S}(\mathcal{F},  ( f + \varepsilon(\cos{\theta}, \sin{\theta}))  - \mathcal{S}(\mathcal{F},  f)\right]} = (0,0).$$
\end{quote}

\subsection{Example.} Let us now consider an example: suppose the feasibility set is (see Fig. 6)
$$\mathcal{F} = \left\{(x,y) \in \mathbb{R}^2:  x \leq 1 \wedge 0 \leq y \leq 1-\frac{x}{2}\right\}$$
and the disagreement pointis  $c=(0.2, 0.1)$. The Nash solution is easily computed to be
$$ \mathcal{S}_{Nash}(\mathcal{F},c) = \left(1, \frac{1}{2} \right).$$
We emphasize that this is an example, where the Nash solution could be considered to be unfair towards Player 2. For comparison, the Kalai-Smorodinsky solution moderately rewards Player 2 for having a slightly higher payoff in the case of disagreement and gives
$$ \mathcal{S}_{KS}(\mathcal{F},c) = \left( \frac{22}{30}, \frac{19}{30} \right) = (0.733, 0.633).$$
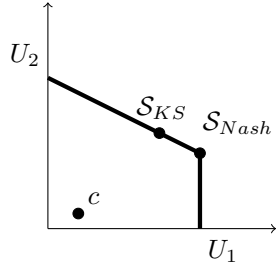
\begin{figure}[h!]
\begin{tikzpicture}[scale = 1]
\draw [->] (0,0) -- (3,0);
\draw [ ->] (0,0) -- (0,3);
\filldraw (0.4,0.2) circle (2pt);
\node at (0.6,0.4) {$c$};
\node at (2.3,-0.3) {$U_1$};
\node at (-0.3,2.3) {$U_2$};
\draw [ultra thick] (0,2) -- (2,1);
\draw [ultra thick] (2,0) -- (2,1);
\filldraw (2,1) circle (2pt);
\filldraw (22/15,19/15) circle (2pt);
\node at (2.5,1.4) {$\mathcal{S}_{Nash}$};
\node at (1.5,1.6) {$\mathcal{S}_{KS}$};
\end{tikzpicture}
\caption{An explicit example where $\mathcal{S}_{Nash}$ could be considered to be unfair.}
\end{figure}

A simple computation yields that if we move the disagreement point to a randomly selected point at distance $\varepsilon$, then the expected payoff for
the Nash solution would be
$$ \mathbb{E} ~ \mathcal{S}_{Nash}(\mathcal{F},c + \varepsilon(\cos{\theta}, \sin{\theta})) \sim \left(1-0.35\varepsilon, \frac{1}{2} + 0.17\varepsilon\right).$$
This means that Player 1 has no incentive to refer to a third party to introduce a random assignment, since he will lose on average. However, Player 2 has a very strong incentive to do so. We emphasize that in a real life scenario, there is no reason to introduce a third party only once and it can usually be repeated an indefinite number of times (the vase, the cutlery, the complete works of Tolstoi, \dots). Analogously, we can compute that for the Kalai-Smorodinsky solution
$$ \mathbb{E} ~ \mathcal{S}_{KS}(\mathcal{F},c + \varepsilon(\cos{\theta}, \sin{\theta})) \sim \left(\frac{22}{30}-0.115\varepsilon^2, \frac{19}{30} + 0.057\varepsilon^2\right).$$
The same phenomenon can therefore be observed for the Kalai-Smorodinsky solution: Player 2 is again being encouraged to have the disagreement point moved.
We should point out that, even though Player 2 is only winning in expectation, our setup allows in principle for Player 2 to have the point moved an arbitrary number of
times which then leads to Player 2 improving his/her situation almost surely.

\subsection{Behavior at the boundary} 
We have yet to address a minor technical point. How do we deal with the situation, where the random process
$$ \mbox{moves}~c~\mbox{to}~c + \varepsilon(\cos{\theta}, \sin{\theta}) \not\in \mathcal{F}?$$
If this were to happen, we would be unable to make sense of the solution at this new point.
\begin{figure}[h!]
\begin{tikzpicture}[scale = 1]

\node at (0.85,0.25) {$c$};
\filldraw (1,0.5) circle (2pt);
\node at (0.85+0.85,0.25+1.25) {$c+ \varepsilon(\cos{\theta}, \sin{\theta}) $};
\filldraw (1.9,1.2) circle (2pt);
\draw [thick] (1,0.5) -- (1.9,1.2);
\draw[ultra thick] (0,1) to[out=0,in=120] (3,0);

\node at (5.85,0.25) {$c$};
\filldraw (6,0.5) circle (2pt);
\node at (5.85+0.85,0.25+1.35) {$c+ \varepsilon(\cos{\theta}, \sin{\theta}) $};
\filldraw (7.1,1.3) circle (2pt);
\draw [thick] (6,0.5) -- (7.1,1.3);
\draw[ultra thick] (5,0.2) to[out=35,in=120] (8,0.31);
\draw [thick,dashed] (6.5,0.85) -- (7.2,0.7);
\filldraw (7.2,0.7) circle (2pt);
\end{tikzpicture}
\caption{Behavior of the process at the boundary: crossing pareto-optimal boundary (left) and crossing at non-pareto boundary (right)}
\end{figure}
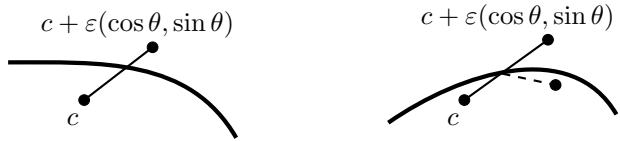
Luckily, there is again a canonical way how to proceed:
if the straight line connecting the old disagreement point with the new disagreement point crosses the pareto-optimal
boundary then that intersection is defined to be the final outcome of the bargaining problem and no further appeal to have the disagreement point moved is possible. Should the intersection be a point
that is not pareto-optimal, we simply reflect around the boundary. We note that prescribing reflection has the side effect of eliminating a list of unnatural solutions
such as, for example, the trivial solution 
$$\mathcal{S}_{triv}(\mathcal{F},c) = c.$$

\section{Characterizing stable solutions}
We now describe our main result. It states that requiring Axiom 2 (\textsc{Invariance under Affine Transformations}) and Axiom 6 (\textsc{Invariance in case of sectoral conflict}) uniquely
characterize a solution $\mathcal{S}_{\Delta}$ of the bargaining problem. We incorporate Axiom 2 in a standard way but only phrasing a solution in the case where $c=(0,0)$, $\mathcal{F} \subset
\mathbb{R}^2_+$ and
$$ \left(\max_{(x,y) \in \mathcal{F}}{x}, \max_{(x,y) \in \mathcal{F}}{y}\right) = (1,1)$$
and then propagating the information through use of the invariance under affine transformations to all solutions -- this is very similar to the way the \textsc{Kalai-Smorodinsky} acts
as a propagated egalitarian solution.

\begin{theorem} There is a unique solution $\mathcal{S}_{\Delta}$ satisfying Axiom 2 (\textsc{Invariance under Affine Transformations}) and Axiom 6 (\textsc{Invariance in case of sectoral conflict}). 
Moreover, the solution automatically satisfies Axiom 1 (\textsc{Symmetry}). $\mathcal{S}_{\Delta}$. It is not always pareto-optimal but in the convex hull of the set of pareto-optimal elements.
$\mathcal{S}_{\Delta}$ can furthermore be explicitly given as the solution of a partial differential equation.
\end{theorem}
As $\mathcal{S}_{\Delta}$ is not always pareto-optimal, it cannot possible be a reasonable solution and this is not what we claim. Instead, we argue that 

\begin{quote}any reasonable solution to the bargaining problem should always dominate $\mathcal{S}_{\Delta}$.
\end{quote}
 The argument in favor of this strong statement draws on the preceding section. Stability in the given context implies that neither player has an incentive to actually refer to a third party. Hence,this additional condition improves the real-world plausibility and robustness of standard results to the bargaining problem.
Conversely, whenever a bargaining solution does \textit{not} dominate $\mathcal{S}_{\Delta}$, then it is in the interest of one of the players to challenge it by appealing to an external court.

To further characterize $\mathcal{S}_{\Delta}$ we suggest to extend our previous argument by allowing for the possibility of iterative sectoral conflicts (first about the vase, then about the house...) for the whole set of items contained in the bargaining set to derive the expected outcome of $\mathcal{S}_{\Delta}$. Practically, this implies the iterative conflicts are brought to court until a pareto-optimal payoff is reached. In what follows we argue, that the average outcome of this process is of natural interest to players as it serves as a vantage point for judging the suitability of a random mode of conflict resolution, which can be achieved by repeatedly appealing to courts. This latter option then serves as an alternative to classical bargaining. As has already been emphasized we suppose that the stability against such conceptual alternatives are a relevant criterion when addressing issues of bargaining.

To simplify considerations at the boundary, we formulate a more symmetric bargaining problem, where every element of the boundary is automatically pareto-optimal (this is accomplished by reinterpreting
the outcome through the use of absolute values). We assume that $\mathcal{F}$ is strictly contained in the positive quadrant $\left\{(x,y): x \geq 0 \wedge y \geq 0\right\}$ and that $c=(0,0)$ and we replace the domain by reflecting the domain into all 4 quadrants and agree that the payoff at $(x,y) \in \mathbb{R}^2$ is understood to be $(|x|, |y|)$. Then this process of iterative conflict, as $\varepsilon \rightarrow 0$, corresponds to a random walk in the domain that goes on until it hits the pareto optimal boundary for the first time.

\begin{figure}[h!]
\begin{tikzpicture}[scale = 1.5]
\draw [ultra thick, ->] (0,0) -- (0,1.5);
\draw [ultra thick, ->] (0,0) -- (1.5,0);
\node at (1, -0.35) {\Large $U_1$};
\node at (-0.35, 1) {\Large $U_2$};
\draw[scale=1,domain=0:1,smooth,variable=\x, ultra thick] plot ({\x},{1-\x*\x});

\draw [ultra thick, ->] (4,-1.5) -- (4,1.5);
\draw [ultra thick, ->] (2.5,0) -- (5.5,0);
\node at (5.4, -0.4) {\Large $U_1$};
\node at (4-0.5, 1.2) {\Large $U_2$};
\draw[scale=1,domain=4:5,smooth,variable=\x, ultra thick] plot ({\x},{1-(\x-4)*(\x-4)});
\draw[scale=1,domain=4:5,smooth,variable=\x, ultra thick] plot ({\x},{-1+(\x-4)*(\x-4)});
\draw[scale=1,domain=3:4,smooth,variable=\x, ultra thick] plot ({\x},{-1+(4-\x)*(4-\x)});
\draw[scale=1,domain=3:4,smooth,variable=\x, ultra thick] plot ({\x},{1-(4-\x)*(4-\x)});

\draw[scale=1,domain=7:8,smooth,variable=\x, ultra thick] plot ({\x},{1-(\x-7)*(\x-7)});
\draw[scale=1,domain=7:8,smooth,variable=\x, ultra thick] plot ({\x},{-1+(\x-7)*(\x-7)});
\draw[scale=1,domain=6:7,smooth,variable=\x, ultra thick] plot ({\x},{-1+(7-\x)*(7-\x)});
\draw[scale=1,domain=6:7,smooth,variable=\x, ultra thick] plot ({\x},{1-(7-\x)*(7-\x)});

\draw [thick] (7,0) -- (7.17,-0.4);
\filldraw (7,0) circle (1pt);
\filldraw (7.17,-0.4) circle (1pt);
\draw [thick] (7.34,0.05) -- (7.17,-0.4);
\filldraw (7.34,0.05) circle (1pt);
\draw [thick] (7.34,0.05) -- (7-0.05,-0.15);
\filldraw (7-0.05,-0.15) circle (1pt);
\filldraw (7-0.26,-0.33) circle (1pt);
\filldraw (7-0.35,-0.75) circle (1pt);
\filldraw (7-0.48,-0.7) circle (1pt);
\draw [thick] (7-0.26,-0.33) -- (7-0.05,-0.15);
\draw [thick] (7-0.26,-0.33) -- (7-0.35,-0.75);
\draw [thick] (7-0.48,-0.7) -- (7-0.35,-0.75);
\end{tikzpicture}
\caption{Copying $\mathcal{F}$ and a sample random walk moving $c$ to the point $(-0.48, -0.77)$. The outcome assigned to the players in this particular event is $(0.48, 0.77)$. The players can expect an average outcome of $\sim (0.59, 0.59)$.}
\end{figure}
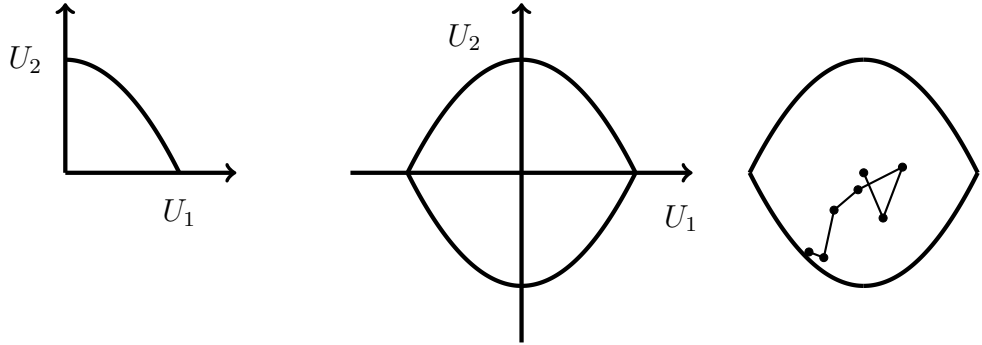
That hitting point then corresponds to the payoff of iterative conflict. The connection between $\mathcal{S}_{\Delta}$ and this random walk is as follows.

\begin{theorem} The expected outcome of iterative conflict is given by $\mathcal{S}_{\Delta}$. 
\end{theorem}
We believe this to be a very strong argument in favor of considering $\mathcal{S}_{\Delta}$ as a natural
quantity in the bargaining problem: surely any 'fair' division should assign to each player a payoff that
is at least as big as can be expect in this fully random scheme that can be enforced by refusing to
cooperate. 

\subsection{Examples} 

 Let us consider two explicit examples. Consider first the feasibility set 
$$\mathcal{F} = \left\{(x,y) \in \mathbb{R}^2: 0 \leq x \leq 1 \wedge 0 \leq y \leq 1-\frac{x}{2}\right\}$$
and the point of disagreement $c = (0,0)$. As mentioned before, this example is typical for one
where the Nash solution punishes Player 2 rather disproportionally.
\begin{figure}[h!]
\begin{tikzpicture}[scale = 0.8]
\draw [ultra thick, ->] (0,0) -- (3,0);
\draw [ultra thick, ->] (0,0) -- (0,3);
\filldraw (0,0) circle (2pt);
\node at (-0.3,-0.1) {$c$};
\node at (2.3,-0.3) {$U_1$};
\node at (-0.3,2.3) {$U_2$};
\draw [ultra thick] (0,2) -- (2,1);
\draw [ultra thick] (2,0) -- (2,1);
\filldraw (2,1) circle (2pt);
\node at (2.5,1.4) {$\mathcal{S}_{Nash}$};
\filldraw (4/3,4/3) circle (2pt);
\node at (4/3+0.1,4/3+0.3) {$S_{KS}$};
\draw [ultra thick] (5,3) -- (7,2);
\draw [ultra thick] (7,2) -- (7,0);
\draw [ultra thick, ->] (5,0) -- (8,0);
\node at (7.5,-0.3) {$U_1$};
\filldraw (7,2) circle (2pt);
\node at (7.5,2.4) {$\mathcal{S}_{Nash}$};
\filldraw (7-4/3,8/3) circle (2pt);
\node at (7-4/3+0.1,8/3+0.3) {$S_{KS}$};
\filldraw (7-2*0.8,2*1.26) circle (2pt);
\node at (7-2*0.8-0.3,2*1.26-0.34) {$S_{\Delta}$};
\end{tikzpicture}
\caption{A bargaining problem, the two classical solutions (left) and the new solution $S_{\Delta}$ in a rescaled
picture (right).}
\end{figure}
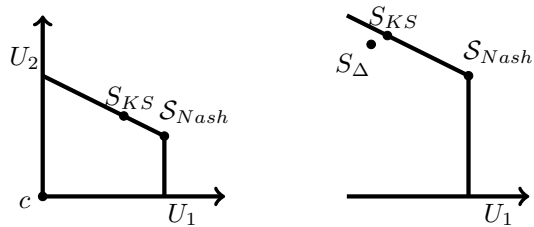
The solutions are
$$ \mathcal{S}_{Nash}(\mathcal{F}, c) = \left(1, \frac{1}{2}\right),
\mathcal{S}_{KS}(\mathcal{F}, c) = \left(\frac{2}{3}, \frac{2}{3}\right) \quad \mbox{and} \quad
 \mathcal{S}_{\Delta}(\mathcal{F}, c) \sim (0.6, 0.63).$$ 
Again, for emphasis, we do not believe $(0.6, 0.63)$ to be a meaningful solution to the problem because it is
not pareto-optimal. Following the reasoning above, however, it does seem reasonable to assign each player
a payoff of at least that size for the simple reason that in the scenario of iterative conflict that's the average of what they can enforce.
Note that the Kalai-Smorodinsky solution indeed dominates $ \mathcal{S}_{\Delta}(\mathcal{F}, c)$ while
the Nash solution does not (echoing the earlier mentioned fact that the Nash solution favors Player 1 
disproportionately). \\

Let us consider a second example favoring Player 2: we take the feasibility set 
$$\mathcal{F} = \left\{(x,y) \in \mathbb{R}^2: 0 \leq x \leq 1 \wedge 0 \leq y \leq 1-x^2\right\}$$
and the point of disagreement $c = (0,0)$. The situation favors the second player and 
$$ \mathcal{S}_{Nash}(\mathcal{F}, c)  \sim (0.57, 0.66), \quad
\mathcal{S}_{KS}(\mathcal{F}, c) \sim \left(0.61, 0.61\right) \quad \mbox{and} \quad
 \mathcal{S}_{\Delta}(\mathcal{F}, c) \sim (0.59, 0.59).$$
This example is curious insofar as the feasibility set is not symmetric but certainly very close to a symmetric set. The
solution $\mathcal{S}_{\Delta}$ is relatively close to the Kalai-Smorodinsky solution, is strongly dominated by it
and, in contrast, not dominated by the Nash solution. We finally consider the completely symmetric problem
$$\mathcal{F} = \left\{(x,y) \in \mathbb{R}^2: x + y \leq 1\right\}$$
with $ c=(0,0)$. Naturally, both the Nash and the Kalai-Smorodinsky solution coincide
because they are both symmetric and pareto-optimal.
$$ \mathcal{S}_{Nash}(\mathcal{F}, c)  = (0.5, 0.5) = \mathcal{S}_{KS}(\mathcal{F}, c).$$
Since $\mathcal{S}_{\Delta}$ is symmetric and contained in the convex hull of the pareto-optimal
part of the boundary (which, in this case, coincides with boundary itself), this immediately implies that we have
$$ \mathcal{S}_{\Delta}(\mathcal{F}, c)  = (0.5, 0.5) \qquad \mbox{as well.}$$

\section{Proof of the theorems}
\subsection{Proof of Theorem 1.} We start by discussing the precise implications of Axiom 6 (\textsc{ISC}) and show that one can actually easily characterize all solutions satisfying Axiom 6.
Let us write 
$$  \mathcal{S}_{\Delta}(\mathcal{F}, c) = (\phi_1(\mathcal{F},c), \phi_2(\mathcal{F},c)).$$
Then Axiom 6 holds if and only if
$$ \left(\frac{\partial^2}{\partial x^2} +  \frac{\partial^2}{\partial y^2}\right) \phi_i(x,y) = 0 \qquad \mbox{for}~i=1,2.$$
Such functions satisfy the so-called mean value property (see e.g. the book of Evans \cite{evans} or Gilbarg \& Trudinger \cite{gil}): for any disk $B(x,r) \subset \mathcal{F}$ of radius $r$ centered around $x$
$$ \frac{1}{2 r \pi}\int_{\partial B(x,r)}{\phi_1(z)d\sigma} = \phi(x).$$
Conversely, a simple Taylor expansion up to second order suffices to show that the partial differential equation
has to be satisfied whenever a Taylor expansion up to second order exists. It is well known (see, for example, Evans \cite{evans})
that functions satisfying that mean-value property necessarily need to be $C^{\infty}$.
The calculation assumes, of course, that $B(c,\varepsilon) \subset \mathcal{F}.$ This characterizes the solution uniquely at
the interior provided we are equipped with conditions at the boundary. It suffices to remark that our characterization of the
random process above immediately implies Dirichlet boundary conditions for the pareto-optimal subset of the boundary
while the reflection implies Neumann conditions on the remainder of the boundary. This partial differential equation is not
invariant under affine transformations: we therefore factor out the symmetries of the group of affine transformations and ask
that henceforth $c=(0,0)$ and 
$$  \left(\max_{(x,y) \in \mathcal{F}}{x}, \max_{(x,y) \in \mathcal{F}}{y}\right) = (1,1).$$
The solution may then be propagated to all domains (this process is, of course, akin to generating the Kalai-Smorodinsky
solution as a propagation of the egalitarian solution via affine transformations).
Altogether, we can thus introduce a function $\phi_1: \mathcal{F} \rightarrow \mathbb{R}$ as the unique solution of the partial differential equation
\begin{align*}
\left(\frac{\partial^2}{\partial x^2} + 
 \frac{\partial^2}{\partial y^2}\right)
 \phi_1(x,y) &= 0 \qquad \mbox{in}~\mathcal{F} \\
\phi_1(x,y) &= x \qquad \mbox{on the strong pareto-optimal part of}~\partial \mathcal{F} \\
\frac{\partial}{\partial \nu} \phi_1(x,y) &= 0 \qquad \mbox{on the remaining subset}~\partial \mathcal{F},
\end{align*}
where $\partial / \partial \nu$ denotes the derivative in tangential direction. Completely analogously, we define a second function $\phi_2: \mathcal{F} \rightarrow \mathbb{R}$
\begin{align*}
\left(\frac{\partial^2}{\partial x^2} + 
 \frac{\partial^2}{\partial y^2}\right)
 \phi_2(x,y) &= 0 \qquad \mbox{in}~\mathcal{F} \\
\phi_2(x,y) &= y \qquad \mbox{on the strong pareto-optimal part of}~\partial \mathcal{F} \\
\frac{\partial}{\partial \nu} \phi_2(x,y) &= 0 \qquad \mbox{on the remaining subset}~\partial \mathcal{F},
\end{align*}
It remains to show that
$$ \mathcal{S}_{\Delta}(\mathcal{F}, c) \qquad \mbox{is in the convex hull of the pareto-optimal boundary.}$$
This follows from the existence
of a harmonic measure and the convexity of the boundary. 
Recall that given the above partial differential equation with
boundary data $g$
we may write the solution as a weighted average
$$ \phi(x,y) = \int_{\partial \mathcal{F}}{g ~d\omega_{x,y}},$$
where $\omega_{x,y}$ is the harmonic measure associated to the point $(x,y)$. By construction, the harmonic measure is supported in the strong pareto-optimal part of
$\partial \mathcal{F}$. Furthermore, the measure is nonnegative and therefore 
$$ (\phi_1(c), \phi_2(c)) = \left(\int_{\partial \mathcal{F}}{x ~d\omega_{c}}, \int_{\partial \mathcal{F}}{y ~d\omega_{c}}\right)$$
and the statement follows immediately.
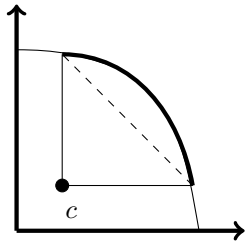
\begin{figure}[h!]
\begin{tikzpicture}[scale = 1.2]
\draw [ultra thick, ->] (4,0) -- (6.5,0);
\draw [ultra thick, ->] (4,0) -- (4,2.5);
\filldraw (4.5,0.5) circle (2pt);
\node at (4.6,0.2) {$c$};
\draw  (4.5,0.5) -- (5.93,0.5);
\draw  (4.5,0.5) -- (4.5,1.95);
\draw (4,2) to[out=0,in=160] (5.1,1.8) to [out=330,in=100] (6,0);
\draw[ultra thick] (4.5,1.95) to [out=360,in=100] (5.93,0.5);
\draw [dashed]  (5.93,0.5) -- (4.5,1.95);
\end{tikzpicture}
\caption{A weighted average over the boundary of a convex set is contained in the convex set. If the curvature is
locally small, the average is very close to the boundary.}
\end{figure}

\subsection{Proof of Theorem 2.} 
Iterating appeal to have the disagreement point moved in a random direction corresponds, via the central limit theorem, to starting a Brownian motion 
in $c$ and letting it wander around until it hits a pareto-optimal point on the boundary. The connection 
to the partial differential equations defining $\phi_1, \phi_2$ is completely standard and follows immediately from
 the Feynman-Kac formula linking harmonic functions to Brownian motion (see e.g. the book of Taylor \cite{tay}).

\section{open problems and remarks}

\subsection{A Pareto-Optimal Solution.} Although $\mathcal{S}_{\Delta}$ is not pareto-optimal, it can be used does to deriveå a natural pareto-optimal solution of the bargaining problem.
Suppose we are given $\mathcal{S}_{\Delta}$ and both players accept Axiom 6 (\textsc{ISC}). Then both players agree that either
of them should be entitled to at least a payoff of $\mathcal{S}_{\Delta}$. Having no further information to go on, it seems reasonable
that both players would agree to replace
$$ (\mathcal{F}, c) \qquad \mbox{by} \qquad (\mathcal{F}, \mathcal{S}_{\Delta}(\mathcal{F},c)).$$ 
Being faced with this new bargaining problem, both will agree that either of them is entitled to a payoff of size at least
$ \mathcal{S}_{\Delta}(\mathcal{F}, \mathcal{S}_{\Delta}(\mathcal{F},c)))$
and therefore agree to replace the bargaining problem by
$$(\mathcal{F}, \mathcal{S}_{\Delta}(\mathcal{F}, \mathcal{S}_{\Delta}(\mathcal{F},c))).$$
Since $\mathcal{S}_{\Delta}(\mathcal{F}, c)$ is a linear combination of pareto-optimal boundary elements, it follows
that this procedure converges to a pareto-optimal element in $\mathcal{F}$. It is not difficult to show that this
happens at exponential speed in the number of iterations.

\subsection{The role of Monotonicity} 
It is currently unclear whether our proposed solution $\mathcal{S}_{\Delta}$ satisfies Axiom 5 (\textsc{Monotonicity}).
This question is of interest, although we have already seen that $\mathcal{S}_{\Delta}$ is uniquely characterized by satisfying Axiom 2 (\textsc{Invariance under Affine Transformations}) and Axiom 6 (\textsc{Invariance in case of sectoral conflict}). In fact, Axiom 2 and Axiom 6 imply that Axiom 1 (\textsc{Symmetry}) is automatically satisfied. In such a case, we could have illustrate the role of Axiom 6 in the context of the traditional solutions in the following way.
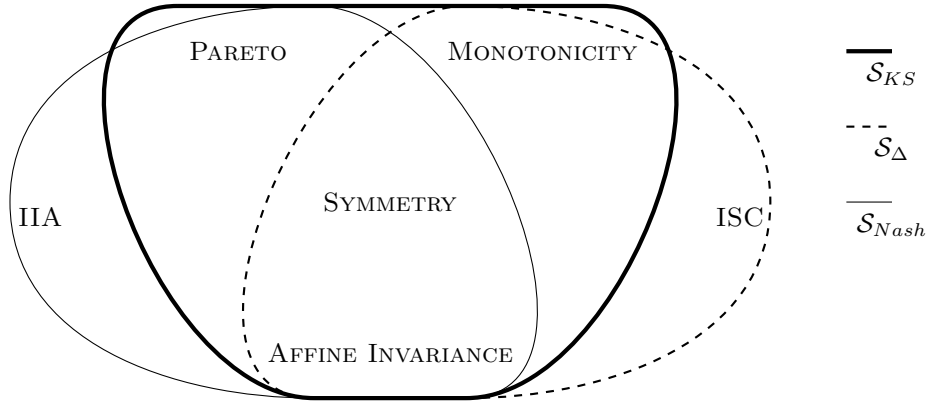
\begin{figure}[h!]
\begin{tikzpicture}[scale = 2]
\node at (-1,1) {$\textsc{Pareto}$};
\node at (1,1) {$\textsc{Monotonicity}$};
\node at (0,0) {$\textsc{Symmetry}$};
\node at (0,-1) {$\textsc{Affine Invariance}$};
\node at (2.3,-0.1) {$\textsc{ISC}$};
\node at (-2.3,-0.1) {$\textsc{IIA}$};
\draw[ultra thick] (-1.4,1.3) to [out=180,in=180] (-0.5,-1.3) to [out=0,in=180] (0.5,-1.3) to [out=0,in=0] (1.4,1.3) to [out=180,in=0] (-1.4,1.3) ;
\draw[dashed, thick] (0.5,1.3) to [out=180,in=180] (-0.5,-1.3) to [out=0,in=180] (0.5,-1.3) to [out=0,in=270] (2.5,0) to [out=90,in=0] (0.5,1.3) ;
\draw[] (-0.5,1.3) to [out=0,in=0] (0.5,-1.3) to [out=180,in=0] (-0.5,-1.3) to [out=180,in=270] (-2.5,0) to [out=90,in=180] (-0.5,1.3) ;

\draw[ultra thick] (3,1)--(3.3,1);
\node at (3.3,0.85) {$\mathcal{S}_{KS}$};

\draw[dashed, thick] (3,0.5)--(3.3,0.5);
\node at (3.3,0.35) {$\mathcal{S}_{\Delta}$};

\draw (3,0)--(3.3,0);
\node at (3.3,-0.15) {$\mathcal{S}_{Nash}$};
\end{tikzpicture}
\caption{Assuming $\mathcal{S}_{\Delta}$ satisfies Axiom 5, the relationship between $\mathcal{S}_{\Delta}$, $\mathcal{S}_{Nash}$, $\mathcal{S}_{KS}$ would be as depicted.}
\end{figure}

\subsection{Variants of Randomness} In the definition of 
random perturbations in case of sectoral conflict as well as in the definition of Axiom
6 (\textsc{ISC}), we chose the 'random' replacement of $c$ by
$$ c + \varepsilon(\cos{\theta}, \sin{\theta})$$
and $\theta$ is uniformly distributed in $[0,2\pi]$. In this context, we wish to emphasize again, that the central limit theorem implies this model of random replacement can be replaced by any other
model as long as it is radially symmetric (i.e. invariant under rotations): the stability of $\mathcal{S}_{\Delta}$ is therefore not tied to any particular form of moving $c$ in a random direction but is a general phenomenon.

\subsection{$\mathcal{S}_{\Delta}$ and the Kalai-Smorodinsky solution} 
Considering the examples above, it is natural to ask whether the Kalai-Smorodinsky solution always dominates $\mathcal{S}_{\Delta}$, i.e.
$$ \mathcal{S}_{KS} \succeq \mathcal{S}_{\Delta}?$$ While we have been unable to find an example, where this relation does not hold, a proof showing that the statement were true would probably be a powerful additional argument in favor of the Kalai-Smorodinsky solution.

\subsection{Computation of $\mathcal{S}_{\Delta}$.} The computation of  $\mathcal{S}_{\Delta}$ is nontrivial and requires
the solution of an elliptic partial differential equation. We emphasize that the actual equation itself is highly stable and
inherits all the usual stability properties of elliptic equations. All the usual classical ways of solving such equation
(i.e. finite elements or the fast multipole method) apply. We also remark that Theorem 2 justifies the use of a Monte
Carlo method based on random walks (which is very easy to implement but inherits the usual slow convergence
speed of Monte Carlo methods).

\section{Outlook} 
\subsection{Trembling games.} The main result of our paper may be summarized as follows.
\begin{quote}
The classical bargaining problem has a unique nontrivial solution $S_{\Delta}$ that is stable under randomly perturbing the disagreement point $c$.
Put differently, other solutions will, in generic situations, not have that property and will be susceptible to small random perturbations
and, in turn, create incentives for one player to try to destabilize the parameters of the game (even though the player has no control
over the direction of the random perturbation).
\end{quote}
It is clear that this underlying idea may have further applications in other areas of game theory; we also emphasize that the underlying
philosophical notion of stability under perturbation is not new, see e.g. Reinhard Selten's \cite{selt} notion of \textit{trembling-hand-perfect} equilibria,
but seems to be novel for the bargaining problem; another structural difference to Selten's notion is that the uncertainty is not in the
players' choice but in the structural design of the game, i.e. it's rather a form of \textit{trembling game stability}.

\subsection{Areas of instability.} We also believe that the converse statement, the solution map not being stable under random perturbations of the disagreement point $c$
for both the Nash solution and the Kalai-Smorodinsky solution (and any other solution for that matter), to be a curious fact that has
been overlooked and a fact that may have serious implications for the actual implementation in large scale bargaining problems (on small scale
problems, the cost of perturbing the game may outweigh the potential benefits). Moreover, it may be an interesting notion in the analyses
of bargaining problems: in particular, it may be used as a tool to partition the bargaining space depending on where random perturbations
would benefit each player. If $\phi_i(c)$ denotes the payoff assigned to player $i$ given disagreement point $c$, then a random
perturbation of $c$ will benefit Player $i$ if and only if
$$  \left[\left( \frac{\partial}{\partial x^2} + \frac{\partial}{\partial y^2} \right) \phi_i \right](c) > 0 \qquad \mbox{whereas}  \qquad \left[\left( \frac{\partial}{\partial x^2} + \frac{\partial}{\partial y^2} \right) \phi_i \right](c) < 0$$
would mean that the Player $i$ is, in expectation, losing payoff under a random perturbation. We demonstrate this in an example
and consider 
$$\Omega = \left\{(x,y) \in \mathbb{R}^2: 0\leq x \leq 1, 0 \leq y \leq 1-x^2\right\}.$$
If the disagreement point is $(0,0)$, then we see that the game is asymmetric and favors Player 2. Moreover, the Nash solution would
assign the payoff $(0.577, 0.666)$. If we now assume the disagreement point to be merely an element in $\Omega$, $c \in \Omega$,
then we can study the effect of moving it in a random direction. A simple computation yields that
\begin{align*}
\phi_1(c_1, c_2) &= \frac{c_1 + \sqrt{3 + c_1^2 - 3c_2}}{3} \\
\phi_2(c_1, c_2) &= 1- \phi_1(c_1, c_2)^2 = 1 - \frac{1}{9}\left(c_1 + \sqrt{3 + c_1^2 - 3c_2}\right)^2
\end{align*}

\begin{figure}[h!]
\begin{tikzpicture}[scale = 3.5]
\draw [thick, ->] (0,0) -- (0,1.2);
\draw [thick, ->] (0,0) -- (1.2,0);
\draw[scale=1,domain=0:1,smooth,variable=\x, ultra thick] plot ({\x},{1-\x*\x});
\draw [thick] (0,0.25) -- (0.87, 0.25);
\draw[thick] (0,1) to [out=320,in=180] (0.28,0.92);
\node at (-0.5,1) {Region 2};
\draw [thick, ->] (-0.2,1) -- (-0.05,0.99);
\node at (0.5,0.1) {Region 1};
\end{tikzpicture}
\caption{The Nash solution: if the disagreement point $c$ is in Region 1, then Player 1 has an incentive to perturb the game, if the disagreement
point is in Region 2, then Player 2 has an incentive to perturb the game.}
\end{figure}
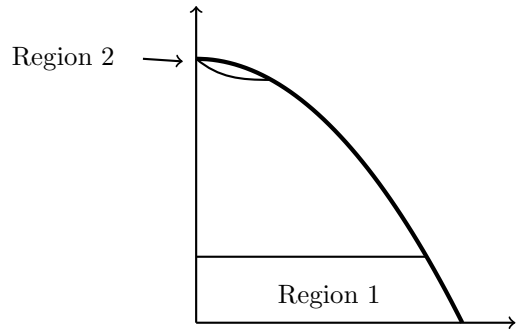

The results are shown in Figure 12. We see the creation of two regions in which the Players have an incentive to perturb the game.
The solution is not stable outside the regions either: here, the Players will, on average, both lose payoff in expectation. We observe 
that this means that Player 1 has an incentive to perturb the game quite a bit (essentially until the disagreement point $c$ has
left Region 1). This seems to pose interesting new challenges: do Regions 1 and 2 ever meet? Does one of them being large
imply that the associated solution is harsh towards that respective player?

\subsection{Conclusion.} Summarizing, we have proposed a new Axiom in the study of the bilateral bargaining problem. The main idea is to introduce
a notion of trembling-hand stability with respect to the disagreement point and to show that this defines a unique solution
satisfying natural properties; the computation of the solution requires solving a partial differential equation. We believe that
the underlying notion could be quite natural in quantifying various asymmetries imposed by a solution on the player and
hope that this will stimulate further research.


\begin{thebibliography}{4}

\bibitem{anbar} D. Anbar and E. Kalai, 
A one-shot bargaining problem. 
Internat. J. Game Theory 7 (1978), no. 1, 13-18. 

\bibitem{anb} N. Anbarci and J. Bigelow, 
The area monotonic solution to the cooperative bargaining problem.
Math. Social Sci. 28 (1994), no. 2, 133-142.

\bibitem{ch} Y. Chun, The equal-loss principle for bargaining problems. Econom. Lett. 26 (1988), no. 2, 103-106. 

\bibitem{evans} L. Evans, Partial differential equations. Graduate Studies in Mathematics, 19. American Mathematical Society, Providence, RI, 1998. 

\bibitem{gil} D. Gilbarg and N. Trudinger, Elliptic partial differential equations of second order. 
Second edition. Grundlehren der Mathematischen Wissenschaften, 224. Springer-Verlag, Berlin, 1983.

\bibitem{ks} E. Kalai and M. Smorodinsky, Other solutions to Nash's bargaining problem. Econometrica 43 (1975), 513--518. 

\bibitem{luce} R. Luce and H. Raiffa, Games and decisions: introduction and critical survey. 
A study of the Behavioral Models Project, Bureau of Applied Social Research, Columbia University; John Wiley \& Sons, Inc., New York, N. Y., 1957.

\bibitem{nash} J. Nash, The bargaining problem. Econometrica 18, (1950). 155--162. 

\bibitem{nash2} J. Nash, Two-person cooperative games. Econometrica 21, (1953). 128-140. 

\bibitem{nic} C. Niculescu and L .E. Persson, Convex Functions and their Applications: A Contemporary Approach. Springer-Verlag. p. 172. ISBN 0-387-24300-3. Zbl 1100.26002.

\bibitem{per} M. Perles and M. Maschler, The super-additive solution for the Nash
bargaining game, International Journal of Game Theory 10 (1981), 163-193.

\bibitem{pe} H. Peters and P. Wakker,
Independence of irrelevant alternatives and revealed group preferences. 
Econometrica 59 (1991), no. 6, 1787-1801. 

\bibitem{selt} R. Selten, Reexamination of the perfectness concept for equilibrium points in extensive games. 
Internat. J. Game Theory 4 (1975), issue 1--2, 25--55. 

\bibitem{sutton} J. Sutton, 
Noncooperative bargaining theory: an introduction. 
Rev. Econom. Stud. 53 (1986), no. 5, 709-724. 

\bibitem{tay} M. Taylor, Partial differential equations. II. Qualitative studies of linear equations. Applied Mathematical Sciences, 116. Springer-Verlag, New York, 1996.

\bibitem{thom} W. Thomson, Bargaining and the theory of cooperative games: John Nash and beyond, Edward Elgar Publishing Ltd, Camberly, Northampton, MA, 2010

\bibitem{yu} P. Yu, “A class of solutions for group decision problems”, Management Science 19 (1973), 936-946.
\end{thebibliography}
\end{document}